\def   \ni {\noindent}

\def   \ssk {\vskip  4truept}
\def   \sk  {\vskip  9truept}
\def   \bsk {\vskip 14truept}

\def   \newline {\hfil\break}

\documentstyle[epsfig]{article}
\begin{document}
\hsize 5truein
\vsize 8truein
\font\abstract=cmr8
\font\keywords=cmr8
\font\caption=cmr8
\font\references=cmr8
\font\text=cmr10
\font\affiliation=cmssi10
\font\author=cmss10
\font\mc=cmss8
\font\title=cmssbx10 scaled\magstep2
\font\alcit=cmti7 scaled\magstephalf
\font\alcin=cmr6 
\font\ita=cmti8
\font\mma=cmr8
\def\ref{\par\noindent\hangindent 15pt}
\null

\title{\ni FIR galaxies and the gamma-ray background}

\bsk \bsk
\author{\ni Andrzej M. So\l tan$^1$ and J\'ozef Juchniewicz$^2$}
\bsk
\affiliation{$^1$ Copernicus Astronomical Center, Bartycka 18, 00-716
Warsaw, Poland} 

\affiliation{$^2$ Space Research Center, Bartycka 18a, 00-716 Warsaw, Poland} \bsk

\baselineskip = 11pt

\abstract{ABSTRACT \ni
Contribution to the gamma-ray background (GRB) by galaxies bright in
the far-IR (FIR) is discussed. Using observational correlations between
the FIR and synchrotron emission in radio wavelengths, it is shown that
the concentration of cosmic ray electrons in FIR luminous galaxies
is substantially higher than in the general population of galaxies.
Microwave background and intrinsic galaxy radiation scattered via the
inverse Compton (IC) effect by cosmic ray electrons into high energies
makes the FIR galaxies relatively strong gamma-ray sources (in comparison
with normal galaxies).  Using the FIR galaxy luminosity function and
the relationship between FIR and radio luminosities, the contribution
(due to the IC only) of FIR luminous galaxies to the GRB above $\sim
1$\,MeV is estimated at $\sim 10$\,\% to $25$\,\% depending on the rate
of cosmic evolution of star burst galaxies.  This is substantially larger
than the normal galaxies contribution. To obtain more accurate assessment
of relationship between the GRB and FIR galaxies, gamma-ray observations
of individual galaxies bright in the FIR are required.}

\ssk
\baselineskip = 12pt

\bsk
\baselineskip = 12pt

\text{\ni 1. INTRODUCTION
\ssk
\ni
The origin of extragalactic diffuse $\gamma$ radiation is unknown (see
Pohl 1998 for a recent review and earlier references). Various
mechanisms producing the $\gamma$-ray background (GRB) have been proposed.
They include rather exotic processes, like annihilation of miscellaneous
particles formed in the early universe (Rudaz \& Stecker 1991) or
evaporation of primordial micro-black-holes (Page \& Hawking 1976),
as well as a superposition of large number of AGNs. However, spectra of
``ordinary'' Seyfert galaxies fall steeply above $\sim 100$\,keV and these
objects contribute mostly to the diffuse background below $\sim 50$\,keV
(Johnson et al. 1994). The most promising candidate producing the GRB
above 10\,MeV seem to be BL Lac objects. Several dozen blazars have been
detected by EGRET and some of them are also strong sources at COMPTEL
energies (von Montigny et al. 1995, Mukherjee et al. 1997).
Quantitative estimates of the blazar contribution to the GRB
rely on the speculative assumptions on the $\gamma$-ray evolution and
vary between $20$ and $90$\,\% (e.g. M\"ucke \& Pohl 1998, Sreekumar et
al. 1997 and references therein). In the present paper we discuss $\gamma$-ray
emission by galaxies luminous in the far-infrared (FIR) domain and the
potential role of these objects for the GRB. We calculate the production of
$\gamma$-rays by cosmic ray electrons scattering low energy photons.
Energetic electrons produce the $\gamma$-rays: {\it a)}
interacting with the interstellar medium (Bremsstrahlung) and {\it b)}
scattering off the cosmic microwave background (CMB) and intrinsic galaxy
radiation \pagebreak via the inverse Compton (IC) process. Although
Skibo \& Ramaty (1993) estimate  that the former process dominates in
the $\gamma$-ray
band, we restrict our investigation to the IC mechanism only. This
is because IC calculations are subject to smaller uncertainties. But
one should note that the total contribution of the FIR galaxies to the
GRB likely exceeds the figures obtained in the paper.

\sk
\ni 2. COSMIC-RAY ELECTRONS IN FIR GALAXIES
\ssk
\ni
The dominant component of radio emission in galaxies is synchrotron radiation
by cosmic ray electrons moving in the magnetic field which fills
the interstellar space. If the electron energy spectrum
has the power law form: $N(E) = N_{\circ} E^{-p}$, using standard formulae
(e.g. Rybicki \& Lightman 1979) one can get:
\begin{equation}
P(\nu) = 1.19\times 10^{-33}\, N_{\circ}\, b^{1.75}\, \nu_9^{-0.75}\,\,
{\rm erg\,s^{-1}Hz^{-1}}\,,
\end{equation}
where $b=B/5\mu{\rm G}$ (B is the strength of magnetic field),
$\nu_9$ is radiation frequency in GHz and we have put $p=2.5$.
Galaxies with a high star formation rate (starburst galaxies) are bright in
the far infrared. They are also rich in cosmic rays, what makes them relatively
bright radio sources. The observational correlation between the luminosities
in these two domains can be fitted using the power law approximation
(Chi \& Wolfendale 1990):
\begin{equation}
\log P_{1.49} = 1.17 \log L_{\rm FIR} + 9.97,
\end{equation}
where $P_{1.49}$ is a luminosity in W\,Hz$^{-1}$ at $\nu = 1.49$\,GHz and
$L_{\rm FIR}$ is the luminosity (in solar units) integrated from 40
to 120\,$\mu$m. Since there is a dependence of $B$ on $L_{\rm FIR}$
(Chi \& Wolfendale 1990)): $B \sim L_{\rm FIR}^{0.125}$, Eq. 1 and 2
could be combined to give the relationship between the $L_{\rm FIR}$
and the normalization of cosmic ray electron distribution:
\begin{equation}
N_{\circ} = 1.67\times 10^{52}\,L_{\rm FIR}^{0.95}\,.
\end{equation}
These cosmic ray electrons produce the $\gamma$-ray photons via the IC
process on the radiation field in the galaxy. In effect, the most luminous
FIR galaxies of $L_{\rm FIR} =10^{12}\,L_{\odot}$ become also strong
$\gamma$-ray sources with luminosities at $10$\,MeV of $L = 3.1\times
10^{38}$erg\,s$^{-1}$MeV$^{-1}$.

\sk
\ni 3. THE GAMMA-RAY BACKGROUND
\ssk
\ni
The total number of cosmic ray electrons in FIR galaxies within 1\,Mpc$^3$
is given by the integral over the FIR luminosity function and the corresponding
local luminosity density at 10\,MeV produced via the IC effect
by these electrons
${\cal L} \approx 1.4\times 10^{40}$\,MeV/(s\,MeV\,Mpc$^3)$.
One can calculate the intensity of the X-ray background (XRB) and GRB produced
by the luminosity density taking into account evolutionary effects:
\begin{equation}
I(E) = {1\over 4\pi}\,{c\over H_{\circ}}\, \int_0^{z_{max}}
{{\cal L}[E(1+z),z] \over (1+z)^2 \sqrt{1+2 q_0 z}} dz,
\end{equation}
\nopagebreak
where ${\cal L}(E,z)$ denotes the luminosity density at energy $E$
at redshift $z$ and all other symbols have their usual meaning.

\begin{figure}
\centerline{\psfig{file=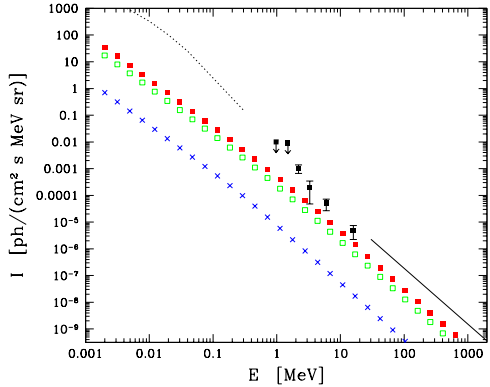, width=7cm}}
\caption{FIGURE 1. Compiled observations of the XRB and GRB: dotted line -
schematic shape adapted from Marshall et al. (1980), full squares with
error bars -
COMPTEL measurements (Kappadath et al. 1995), solid line - EGRET
(Sreekumar et al. 1997). Expected background flux produced
by FIR galaxies: crosses - CMB photons only, no evolution; open squares -
all photons, $\beta = 2.5$; full squares - all photons, $\beta = 3.0$
(see the text for details).} \label{bkg_fit}
\end{figure}

Evolution of the luminosity density ${\cal L}(E,z)$ results from the
dependence of the CMB temperature on redshift, evolution of the FIR
radiation field and the variations of the cosmic electron
concentration in galaxies. In Fig.\,1 we show the predicted background
for selected models. Crosses denote the background produced
by scattering of CMB photons only assuming {\it no evolution} of cosmic ray
electrons. At 10\,MeV the model produces roughly 0.5\,\%
of the observed background. However, it is likely that the population of
galaxies luminous in the FIR is subject to strong evolution. The redshift
distribution of IRAS galaxies (e.g. Saunders et al. 1990) is consistent
with the density evolution $\sim (1+z)^{6.7\pm 2.3}$ for $z < 0.25$.
Samples are not deep enough to provide direct information on the FIR
evolution at higher redshits, but it is generally accepted that
starburst galaxies, which are the major constituent of the FIR
galaxies, exhibit a strong increase of the spatial density and/or
luminosity up to $z \approx 2.5$ with a decay at higher redshifts
(e.g. Moorwood 1996, Franceschini et al. 1997, Bechtold et al. 1998).
In agreement with the available data we have used the following
parametrization of the evolution of the FIR luminosities:
\begin{equation}
L_{\rm FIR}(z) = L_{\rm FIR}(0)~ \left\{ \begin{array}{lcl}
(1+z)^{\beta}                              & {\rm ~~for~~} & z \le z_{\star}\,, \\
(1+z_{\star})^{\beta} \exp{\{-a(z-z_{\star})\}} &          & z > z_{\star}\,.
\end{array}
\right .
\end{equation}
where we put $z_{\star} = 2.5$, $a = 1.25$. We have considerd two rates
of evolution: $\beta = 2.5$ and $3.0$. The XRB and GRB produced by
galaxies subjected to these evolution models is shown in
Fig.\,1 with open squares for $\beta = 2.5$ and with full squares for
$\beta = 3.0$. It is conspicuous that at energies above~ $\sim 5$\,MeV
the model with $\beta = 3.0$ produces 20 - 25\,\% of the observed background.

\bsk
\ni 4. PROSPECTS FOR THE FUTURE}
\ssk
\ni
Indirect methods, which use objects selected at one energy band to
predict the background at another energy and exploit correlations between
luminosities in these two bands are not capable to provide accurate
estimates of the background. Studies of the soft X-ray background in
the last 20 years clearly show that. However, the present calculations
demonstrate that the FIR galaxies potentially are an important constituent
of the GRB. To determine the contribution of these objects to the GRB
one should make direct observations in the $\gamma$-ray band of nearby
IRAS sources -- known bright starburst galaxies. Unfortunately, prospects
for the detection of these galaxies in the $\gamma$-ray range  with the
present day instruments are discouraging. This is because in the present
model a substantial fraction of the GRB is produced by a large number of
relatively weak $\gamma$-ray sources. EGRET observations show that some
AGNs reach luminosities of $10^{47}-10^{49}$\,erg\,s$^{-1}$ (von Montigny
et al. 1995), while in our model local $\gamma$-ray luminosities due to
the IC process cover the range of $10^{36}-10^{39}$\,erg\,s$^{-1}$. The
total $\gamma$-ray luminosities of FIR galaxies are most probably
substantialy larger because of the Bremsstrahlung, but still are several
orders of magnitude below blazar luminosities. However, it is likely that
the relatively high spatial density of FIR galaxies makes them dominant
contributors to the GRB.

\bsk
\baselineskip = 12pt
{\abstract \ni ACKNOWLEDGMENTS
This work has been partially suported by the Polish KBN grant
\break 2\,P03C\,009\,10.}

\ssk
\baselineskip = 12pt

{\references \ni REFERENCES
\ssk
\ref Bechtold J., Elston R., Yee H. K. C., Ellingson E. and Cutri R. M., 1998,
{\it atro-ph/9802230}
\ref Chi X. and Wolfendale A. W., 1990, MNRAS 245, 101
\ref Franceschini A., Aussel H., Bressan A., Cesarsky C. J., Danes L.,
De Zotti G., Elbaz D., Granato G. L., Mazzei P. and Silva L., 1997,
{\it astro-ph/9707080}
\ref Johnson W. N. et al., 1994, {\it 2$^{nd}$ Compton Symposium},
AIP Conf. Proc., 304, 515
\ref Kappadath S. C. et al., 1995, in Proc. 24th Intl. Cosmic-Ray Conf.
(Rome), vol. 2, 230
\ref Marshall R. F., Boldt A., Holt S. S. et al., 1980, ApJ 235, 4
\ref Moorwood A. F. M., 1996, {\it ESO Preprint No. 1170}
\ref M\"ucke A. and Pohl M., 1998, {\it astro-ph/9807297}
\ref Mukherjee R. et al., 1997, Ap0J 490, 116
\ref Page D. N. and Hawking S. W., 1976, ApJ 206, 1
\ref Pohl M., 1998, 16th ECRS GR2 Lecture, {\it astro-ph/9807267}
\ref Rybicki G. B. and Lightman A. P., 1979, {\it Radiative Processes in
Astrophysics}, John Wiley \& Sons, New York
\ref Rudaz S. and Stecker F. W., 1991, ApJ 368, 40
\ref Saunders W., Rowan-Robinson M., Lawrence A., Efstathiou G., Kaiser N.,
Ellis R. S. and Frenk C. S., 1990, MNRAS 242, 318
\ref Skiboa J. G. and Ramaty R., 1993, A\&ASS 97, 145
\ref Sreekumar P., Stecker F. W. and Kappadath S. C., 1997, {\it astro-ph/9709258}
\nopagebreak
\ref von Montigny C. et al., 1995, ApJ 440, 525
}
\end{document}